\documentclass{webofc}
\usepackage[varg]{txfonts}   
\usepackage{color}

\usepackage{blindtext}
\usepackage{hyperref}
\hypersetup{
    colorlinks=true,
    allcolors=blue,
    filecolor=blue,      
    urlcolor=blue,
    }

\begin{document}

\title{CONCERTO at APEX: 
Installation and first phase of on-sky commissioning}
%
%

\author{
\firstname{A.~Catalano}\inst{4,6}\fnsep\thanks{\email{catalano@lpsc.in2p3.fr}}
\and 
\lastname{P.~Ade} \inst{1}
\and 
\lastname{M.~Aravena}\inst{2}
\and 
\lastname{E.~Barria} \inst{3,4}
\and 
\lastname{A.~Beelen} \inst{5}
\and 
\lastname{A.~Benoit} \inst{3,4}
\and 
\lastname{M.~B\'ethermin} \inst{5}
\and 
\lastname{J.~Bounmy} \inst{6,4}  
\and 
\lastname{O.~Bourrion} \inst{6,4}  
\and 
\lastname{G.~Bres} \inst{3,4}
\and 
\lastname{C.~De Breuck} \inst{7}
\and 
\lastname{M.~Calvo} \inst{3,4}
\and 
\lastname{F.-X.~D\'esert} \inst{8,4}
\and 
\lastname{C.A Dur\'an} \inst{9}
\and 
\lastname{G.~Duvauchelle} \inst{5}
\and 
\lastname{L.~Eraud} \inst{6}
\and 
\lastname{A.~Fasano} \inst{3,4}  
\and 
\lastname{T.~Fenouillet} \inst{5}
\and 
\lastname{J.~Garcia}  \inst{5} 
\and 
\lastname{G.~Garde} \inst{3,4} 
\and 
\lastname{J.~Goupy} \inst{3,4}
\and 
\lastname{C.~Groppi} \inst{10}
\and 
\lastname{C.~Hoarau} \inst{6,4}  
\and 
\lastname{W.~Hu} \inst{5}
\and 
\lastname{G.~Lagache} \inst{5}
\and 
\lastname{J.-C.~Lambert} \inst{5}
\and 
\lastname{J.-P.~Leggeri} \inst{3,4}
\and 
\lastname{F.~Levy-Bertrand} \inst{3,4}
\and 
\lastname{J.~Mac\'{\i}as-P\'erez} \inst{6,4}
\and 
\lastname{H.~Mani}\inst{10}
\and 
\lastname{J.~Marpaud} \inst{6,4}
\and 
\lastname{M.~Marton} \inst{6}
\and 
\lastname{P.~Mauskopf} \inst{10}
\and 
\lastname{A.~Monfardini} \inst{3,4}
\and 
\lastname{G.~Pisano} \inst{1}
\and 
\lastname{N.~Ponthieu} \inst{8,4} 
\and 
\lastname{L.~Prieur} \inst{5}
\and 
\lastname{G.~Raffin} \inst{6}
\and 
\lastname{S.~Roni} \inst{6}
\and 
\lastname{S.~Roudier} \inst{6}
\and 
\lastname{D.~Tourres} \inst{6,4}
\and 
\lastname{C.~Tucker} \inst{1}
\and 
\lastname{L.~Vivargent} \inst{6}
}

\institute{
Astronomy Instrumentation Group, University of Cardiff, The Parade, CF24 3AA, United Kindgom
\and
N\'ucleo de Astronom\'ia, Facultad de Ingenier\'ia y Ciencias, Universidad Diego Portales, Av.  Ej\'ercito 441, Santiago, Chile
\and
Univ. Grenoble Alpes, CNRS, Grenoble INP, Institut N\'eel, 38000 Grenoble, France
\and 
Groupement d'Interet Scientifique KID, 38000 Grenoble and 38400 Saint Martin d'H\'eres, France
\and 
Aix Marseille Universit\'e, CNRS, LAM (Laboratoire d'Astrophysique de Marseille), F-13388 Marseille, France
\and 
Univ. Grenoble Alpes, CNRS, LPSC/IN2P3, 38000 Grenoble, France
\and 
European Southern Observatory, Karl Schwarzschild Straße 2, 85748 Garching, Germany
\and
Univ. Grenoble Alpes, CNRS, IPAG, 38400 Saint Martin d'H\'eres, France
\and
European Southern Observatory, Alonso de Cordova 3107, Vitacura, Santiago, Chile
\and
School of Earth and Space Exploration and Department of Physics, Arizona State University, Tempe, AZ 85287, USA
}

\abstract{
  CONCERTO (CarbON CII line in post-rEionisation and ReionisaTiOn) is a large field-of-view (FoV) spectro-imager that has been installed on the Cassegrain Cabin of Atacama Pathfinder EXperiment (APEX) telescope in April 2021. CONCERTO hosts 2 focal planes and a total number of 4000 Kinetic Inductance Detectors (KID), with an instantaneous FoV of 18.6 arc-minutes in the range of 130-310 GHz. The spectral resolution can be easily tuned down to 1\,GHz depending on the scientific target. The scientific program of CONCERTO has many objectives, with two main programs focused on mapping the fluctuations of the [CII] line intensity in the reionisation and post-reionisation epoch (4.5<z<8.5), and on studying galaxy clusters via the thermal and kinetic Sunyaev-Zel'dovich (SZ) effect. CONCERTO will also measure the dust and molecular gas contents of local and intermediate-redshift galaxies,  it will study the Galactic star-forming clouds and finally it will observe the CO intensity fluctuations arising from 0.3<z<2 galaxies.
  
  The design of the instrument, installation at APEX and current status of the commissioning phase and science verification will be presented. Also we describe the deployment and first on-sky tests performed between April and June 2021.
}
\maketitle
\section{Introduction}
\label{intro}

Starting from January 2019, the Grenoble collaboration (LPSC\footnote{\emph{Laboratoire de Physique Subatomique et Cosmologie}}, Institut N\'eel and IPAG \footnote{\emph{Institut de plan\'etologie et d'astrophysique de Grenoble}}) together with the LAM\footnote{\emph{Laboratoire d'Astrophysique de Marseille}}, funded by an Advanced Grant ERC has worked to design, fabricate, install, commission, and observe with the CONCERTO \cite{concerto2}  instrument. CONCERTO is installed at the 12-metre APEX telescope and now it is open to the APEX community, the scientific goals will be manifold. The CONCERTO collaboration will focus on two cosmological goals: 

\begin{itemize}
  \item The measurement of variations of the [CII] line emission at redshift z>5.2. The atomic [CII] line is one of the most valuable tracers of star formation. At high redshifts it is observed in the sub-millimetre and millimetre atmospheric windows~\cite{lagache,concerto}. We will use the [CII] line emission as a tracer of cosmic density structure, and make the first constraints on the power spectrum of dusty star-forming matter. Our experiment will also observe the CO intensity fluctuations arising from galaxies at redshifts between 0.2 and 2, mapping the spatial distribution and abundance of molecular gas over a broad range of cosmic time.



 \item Precise measurements of the shape of the SZ electromagnetic spectrum for cluster of galaxies for redshifts between 0.2 and 0.8 to disentangle the kinetic and relativistic corrections from the main thermal contribution~\cite{concerto}. 

\end{itemize}

Since April 2021 CONCERTO is installed at the 12 metre APEX telescope, located at a 5105~m altitude on the Llano de Chajnantor in Northern Chile \cite{apex}. 
The commissioning phase has been performed until the end of June. Starting from July 2021 the instrument has started its regular scientific program. In the next sections we will present a description of the instrument, its installation at the telescope and some technical and scientific commissioning results. 

\begin{figure}
\centering
\includegraphics[width=12cm,angle=0]{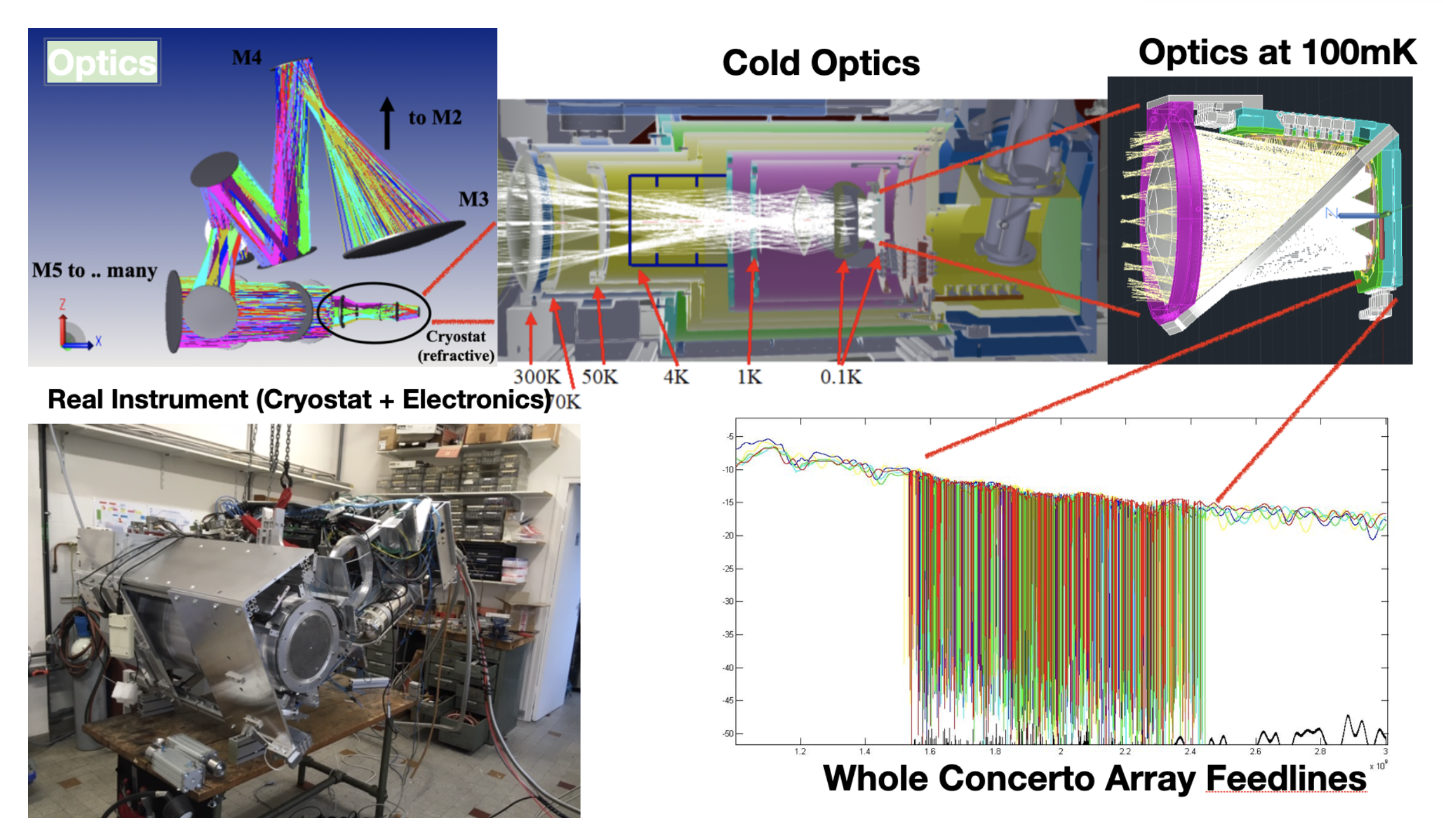}
\caption{Top: CONCERTO optical design (using Zemax software with details of the cold optics inside the cryostat and the 100\,mK stage cold pupil, polariser and the two arrays). Bottom left: picture of the fabricated CONCERTO cryostat and warm electronics in the laboratory at Neel in Grenoble. Bottom right: baseline sweeps for one array (6 feedlines) taken with Vector Network Analyser (VNA).}
\label{concerto_instrument}
\end{figure}

\section{The CONCERTO Instrument}
\label{sec-1}

The instrument is based on several sub-systems: an optical box which selects the optical inputs to be injected in a second box that contains a Martin-Puplett interferometer. The signal is then focused on two arrays of about 4000 KID detectors, cooled at 100\,mK through a closed-cycle dilution cryostat (see Fig.~\ref{concerto_instrument}). 

The design and the control of the systematic errors of such a configuration drives the major requirement for the whole instrument. This is because the MPI must be able to perform continuously up to 9\,cm path interferograms, which gives a spectral resolution of about 1\,GHz within the band, at a frequency of about 4\,Hz in order to maintain a low sky noise level. The fast KID time response permits us to use this method without any loss of information. The readout sampling frequency is constrained by this requirement; data are sampled at 4\,kHz producing roughly 128\,MByte per second. 

The optical system has been arranged to have a FoV of 18.6\,arcmin. Since an MPI measures the difference between two optical inputs \cite{mpi}, we split the beam after passing through the telescope into two fully polarised beams before reaching the MPI. This can be done with the use an initial wire-grid polariser. After interfering in the MPI, the first half of the beam is focused onto the detectors, creating an image of 18.6 arcmin on the sky. The second half of the beam is first de-focused from a focused image to a pupil (corresponding to the image of the primary mirror of the telescope) and then collected to the detectors. This makes an initial hardware subtraction of the constant common mode from the atmosphere to simplify the off-line analysis. In addition, instead of the sky itself, the second input can be taken from a cold reference, which is a cold black-body at a given tunable temperature.

The 18.6 arcmin focal plane is fully covered by two arrays of single polarisation aluminium LEKID containing 2152 pixels each and separated by a 45 degrees polariser. Each array has a different optical band 195-310\,GHz for the High frequency array (HF) and 130-270\,GHz for low frequency array (LF).


We recall in Table\,\ref{tab1} the main characteristics of CONCERTO. 

\begin{table}[ht]
{
\begin{center}
\begin{tabular}{c|c}
Telescope primary mirror diameter [m] & 12 \\ \hline
Field-of-view diameter [arcmin] & 18.6 \\\hline
Beam Widths [arcsec] & 30 (HF) 35 (LF) \\\hline
Absolute spectral resolution [GHz] & $\geq$ 1 \\\hline
Relative spectral resolution R [\#] & 1--300 \\\hline
Frequency range HF $\mid$ LF [GHz] & 195--310 $\mid$ 130--270 \\\hline
Pixels on Sky HF $\mid$ LF [\#] & 2,152 $\mid$ 2,152 \\\hline
Instrument geometrical throughput [sr\,m$^2$] & 2.5$\times$10$^{-3}$\\\hline
Single Pixel geometrical throughput [sr$\,$m$^2$] & 1.16$\times$10$^{-6}$\\\hline
Data rate [MBytes/sec] & 128 \\
\end{tabular}
\end{center}
}
\vspace{-.5cm}
\caption{Main characteristics of CONCERTO at APEX.} 
\label{tab1}
\end{table}

\begin{figure}
\centering
\includegraphics[width=12cm,angle=0]{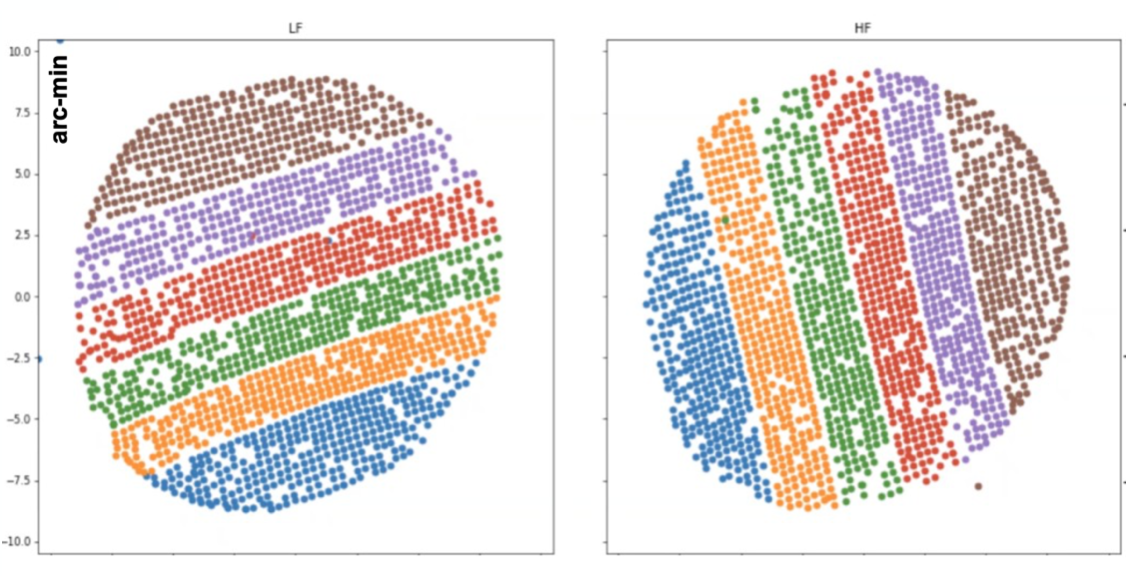}
\caption{LF (left) and HF (right) focal plane geometries. Around 90\% of the designed pixels exhibit a beam and are healthy.}
\label{concerto_beam}
\end{figure}

\section{ Installation and first phase of on-sky commissioning}
\label{sec-2}

After installation of the whole instrument, the cryostat has been first cooled down. After reaching a base temperature of 70\,mK CONCERTO has been switch on for the first time the 12th of April 2021. The first phase of the technical commissioning started right after. During this phase, the sub-systems operations and typical electrical characteristic of the electronics and detectors (feedlines connections, quality factors of the detectors, feedline cosmetics, etc..) were checked. The optical alignment has been achieved using an optical laser and a very first light on Jupiter was obtained. 

After these first checks we started to point the telescope in order to start the scientific commissioning. We list, among others, the following achievements:\\

\begin{itemize}
    \item About 90\% of the detectors are valid and able to detect bright sources individually on the sky. 
    
    \item A fraction of about 70\% of the designed KIDs exhibits a beam with an eccentricity lower than 0.7.
    
    \item A preliminary Noise Equivalent Temperature (NET) has been evaluated and equal to  2.2\,mK/$\sqrt{Hz}$,  which is well in agreement with the expected value. Note that we expect to have a better sensitivity per beam at low frequency where we have more than one KID per beam, and at intermediate frequencies where the two arrays overlap in frequency.

\end{itemize}

\begin{figure}
\centering
\includegraphics[width=7.6cm,angle=0]{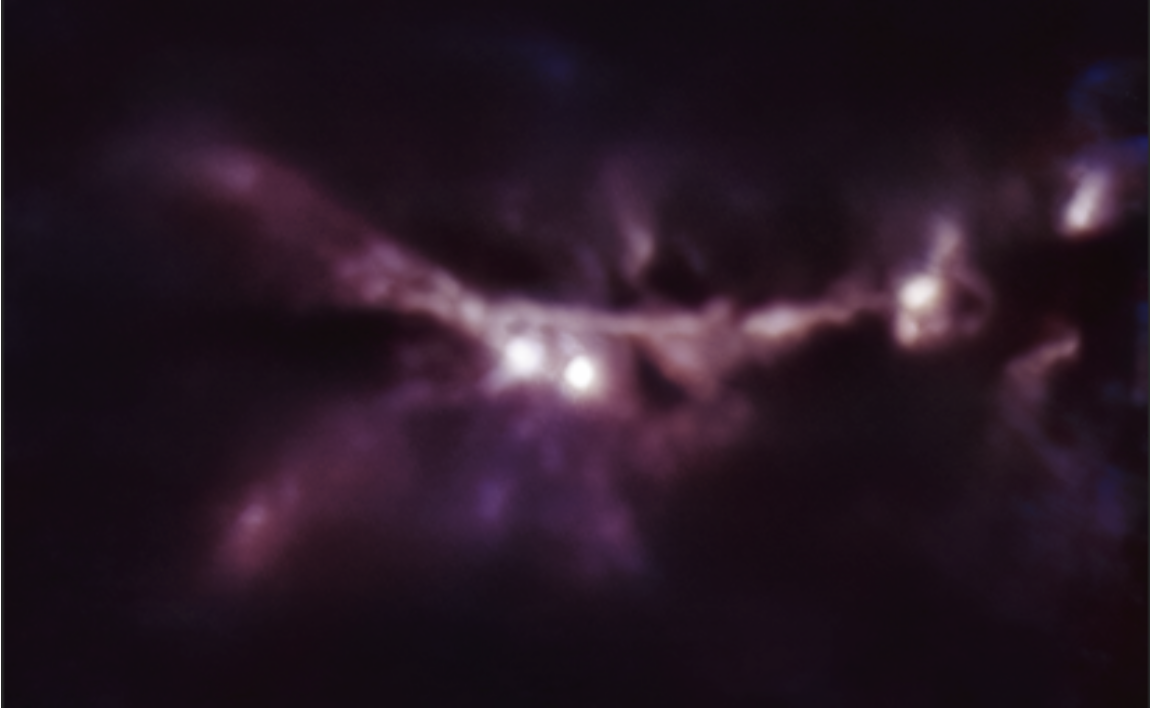}
\includegraphics[width=4.9cm,angle=0]{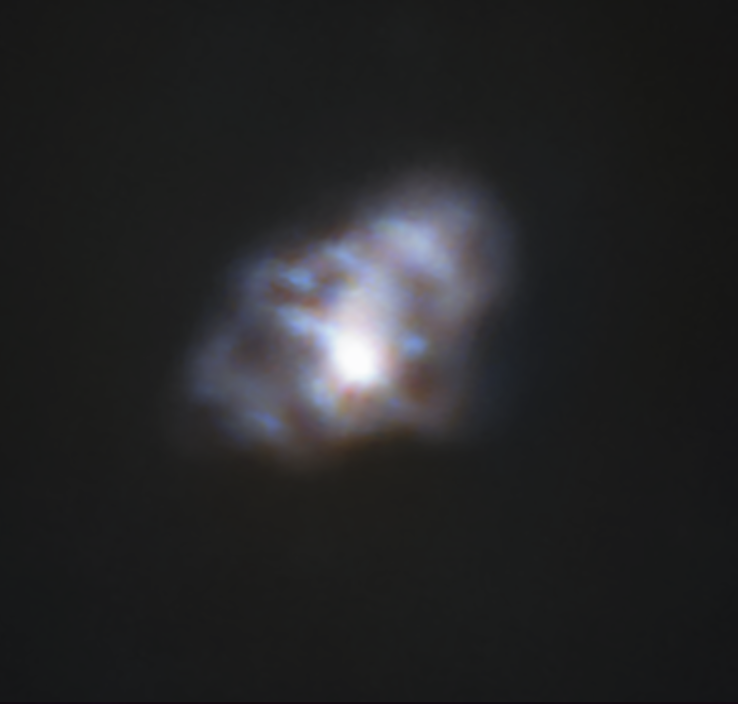}
\caption{First results of the CONCERTO scientific commissioning. Two observations of continuum emission are shown as observed by the two bands of CONCERTO (\ref{tab1}). Left plot: star forming region NGC6334 (Cat Paw Nebula) mapped in a field of 37$\times$25\,arcmin$^2$ for 16\,minutes of integration. Right plot: Crab nebula observed with 2.5 minutes integration time mapped in a field of 10$\times$10\,arcmin$^2$. In both cases we represent composite images where the LF array is in blue, and the HF array in red (ESO/CONCERTO collaboration: \url{https://www.eso.org/public/unitedkingdom/announcements/ann21010/?lang}).}
\label{concerto_photo}
\end{figure}

The presented results must be consolidated with detailed data analysis, which is still in progress. 
Before the end of the scientific commissioning few selected astronomical targets were observed to demonstrate the capabilities of the instrument and optimize the observing strategies. In particular in this paper we present in Figure \ref{concerto_photo} photometric maps (integrated emission on the full CONCERTO band) of the star forming region NGC6334 (the Cat Paw Nebula) and of the Crab nebula supernova remnant. 

\section{Conclusions, Future work and perspectives}

The CONCERTO instrument has been successfully installed on the APEX telescope in April 2021. Operations were started few days after the installation and scientific commissioning was performed in less than two months. 

Here we presented preliminary results of the scientific commissioning showing estimates of the instrument performance well in agreement with expectations. 

The regular scientific observations have already started in July 2021 thanks to the fact that the instrument has a complete remote operations design. Indeed, after installation it became impossible to travel to Chile, therefore operations of the instrument are all conducted remotely from France.

Spectra and further results of the commissioning will be presented in a more extensive paper in the coming months. Scientific commissioning data are available in the ESO data base at \url{ https://www.eso.org/sci/publications/announcements/sciann17437.html.}\\
\\
\\
\small
{\bfseries \emph{Acknowledgements.}} This project has received funding from the European Research Council (ERC) under the European Union’s Horizon 2020 research and innovation programme (grant agreement No 788212) also from the Excellence Initiative of Aix-Marseille University-AMidex, a French "Investissements d’Avenir" programme and the labex FOCUS. We thank the electronics, mechanics, cryogenics and administrative groups at Institut N\'eel, LPSC and LAM. For the outstanding technological and human support we thank the whole APEX staff.

%
%
%

\end{document}